\documentclass[aps, prl, reprint, superscriptaddress]{revtex4-1}
\usepackage{graphicx}%
\usepackage{dcolumn}%
\usepackage{bm}%
\usepackage{amsmath}
\usepackage{amstext}
\usepackage{amssymb}
\usepackage{amsthm}

\usepackage{xcolor}
\usepackage{empheq}
\usepackage{stmaryrd}

\newcommand{\vect}[1]{\mathbf{#1}}
\newcommand{\tens}[1]{\mathsf{#1}}

\DeclareMathOperator{\diag}{diag}

\usepackage[hidelinks]{hyperref}

\renewcommand{\d}{\mathrm{d}}

\newcommand{\1}{\mathbf{1}}

\renewcommand\epsilon{\varepsilon}

\usepackage[normalem]{ulem}

\begin{document}

\title{Elastic instability behind  brittle fracture}

\author{D. Riccobelli}
\affiliation{MOX -- Dipartimento di Matematica, Politecnico di Milano, 20133 Milano, Italy}
\author{P. Ciarletta}
\affiliation{MOX -- Dipartimento di Matematica, Politecnico di Milano, 20133 Milano, Italy}
\author{G. Vitale}
\affiliation{Laboratoire de Mecanique des Solides, École Polytechnique, 91128 Palaiseau, France.}
\author{C. Maurini}
\affiliation{CNRS, Institut Jean Le Rond d’Alembert, Sorbonne Université, UMR 7190, 75005 Paris, France.}
\author{L. Truskinovsky}
\email{lev.truskinovsky@espci.fr}
\affiliation{ESPCI ParisTech, PMMH, CNRS -- UMR 7636, 75005 Paris, France.}

\graphicspath{{figures}}

\begin{abstract}

 We argue that nucleation of brittle cracks in initially flawless soft elastic solids is preceded by a nonlinear elastic instability, which cannot be captured without accounting for  geometrical precise description  of finite elastic deformation.  As a  prototypical problem we consider  a homogeneous  elastic  body subjected to tension and assume that it is weakened by the presence of a free surface which then serves as a site of crack nucleation. We show that in this maximally simplified setting,   brittle fracture emerges from a symmetry breaking elastic instability  activated by   softening  and involving large elastic rotations.   The implied bifurcation of  the homogeneous elastic equilibrium   is highly unconventional for nonlinear elasticity as it exhibits an extraordinary sensitivity to geometry,  reminiscent of  the transition to turbulence in fluids. We trace the post-bifurcational development of this instability beyond the limits of applicability of scale free continuum elasticity and  use  a phase-field approach  to  capture the scale dependent sub-continuum  strain localization,  signaling the formation of actual  cracks.

\end{abstract}

\maketitle

While linearized elasticity theory  is  usually sufficient in problems involving \textit{propagation} of pre-existing cracks \cite{landau1986theory,broek2012practical,broberg1999cracks},
we present a compelling evidence that, at least for some classes of soft materials, the description of crack \textit{nucleation}  requires an account of both geometrically and physical elastic nonlinearity \cite{stuart2nonlinear, Marsden1983}.  To elucidate the physical origin of  the failure of  linear theory,  we  build a continuous path from surface  instability in tension  to fracture.

The phenomenon of \emph{surface fracture} is of considerable recent interest because  the sub-micron  parts employed in many modern  applications  are   effectively defect free  and their   fracture  usually  originates on unconstrained external  surfaces   \cite{dehm2018overview}. Crack nucleation at the surface  is also of importance for the understanding of   the fragmentation of various  brittle surface  layers  \cite{hutchinson1991mixed,colina1993model, handge2000two,alarcon2010softening}. More generally, the emergence of surface fracture patterns  \cite{bourdin2014morphogenesis,truskinovsky2014mechanical} is an example of a symmetry breaking instability which is at the heart of complexity development from   soft  matter physics \cite{kim2010dynamic,Maha2011} to  biophysics   \cite{amar2005growth,ciarletta2014pattern}.

Nonlinear elastic instabilities  were  studied extensively in the context of \emph{compressive}  buckling   \cite{BiotBook, OgdenBook,bigoni2012nonlinear,HH75,DPieroRizz,Spector2012,grabovsky2007flip,SAWYERS1982,Triantafyllidis2007}.
 Elastic  instabilities can also take place under tension, with necking, wrinkling and shear banding, as the most prominent examples \cite{audoly2016analysis,li2016stability,cerda2003geometry,hutchinson1981shear,hill1975bifurcation}.   However, the potential  relation of these \emph{tensile} instabilities to fracture has been largely overlooked.     Several studies  attempted to develop  conceptual links between the  bulk crack nucleation and  material softening and  used  them  to advance   various  phenomenological nucleation criteria \cite{klein1998crack, silling2010crack, bourdin2014morphogenesis, shekhawat2013damage, tanne2018crack, kumar2020revisiting, kristensen2021assessment, hao2019atomistic, larsen2023variational}. Still, an understanding of how such criteria relate to the subtle interplay between geometric and physical nonlinearities along the  crack nucleation path  remains obscure.
 
 Brittle cracking of soft solids is not uncommon, as it is exemplified  by an  abrupt failure of an elastic  rubber band under tension. In particular,  \emph{brittle-soft}  behavior is characteristic for hydrogels  [39–42],  where the diverging stress at the crack tip is typically accompanied not only by   large stretches but also by  large rotations with several candidate mechanisms debated as potential regulators of the underlying material failure at the micro scale [43].

In this Letter we use the geometrically simplest setting to explore  both linear and nonlinear stages of the tensile instability in a soft solid which culminates in the formation of a brittle  crack. The  implied instability is  of  \emph{spinodal} type  \cite{gagne2005simulations,schweiger2007transient, klein2002nucleation}  but with a peculiarity that it is associated with the  surface rather than with the bulk \cite{Simpson_1985, benallal1993bifurcation, simpson2020complementing, mielke1998quasiconvexity, negron2012violation}. The degenerate   nature of this  instability in the purely elastic setting \cite{mikhlin1973spectrum}  leads to a high  sensitivity of the emerging   patterns to sample geometry.  Such sensitivity is typical for nonlinear  systems without an internal  length  scale and therefore the ensuing crack nucleation scenario  is  reminiscent, for instance,  of  a transition to turbulence. Regularization of the problem, bringing  a fixed internal length scale,  naturally simplifies the picture, as it is already known from the study of the prototypical one dimensional models \cite{salman2021localizing,gupta2023nucleation}. 

Our first goal is to   show  in detail how  the above  symmetry breaking elastic instability serves as a precursor of the  ultimate  strain localization. Then, since the emerging strain singularity renders the  scale-free continuum elasticity inadequate,  the modeling paradigm must be   changed if the goal is   to   capture the formation of sharp cracks.  To describe the role of  micro-scales  in such a sharpening process we resort to a    phase-field-type extension of the  continuum theory  \cite{BOURDIN2000797,Karma01,tanne2018crack,salman2021localizing}. We show that such a hybrid approach  allows one to model seamlessly the whole process from a continuum  elastic instability to  a   sub-continuum  evolution of developed cracks.

\begin{figure}[h!]
    \centering
    \includegraphics[width=\columnwidth]{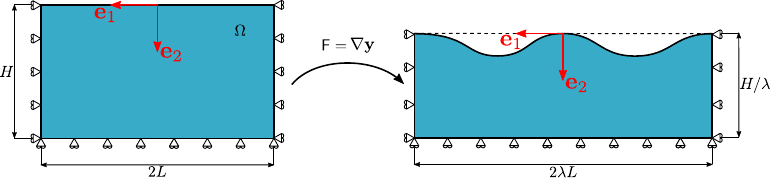}
    \caption{Schematic representation of the considered  surface instability  showing the reference and the actual configurations,  while also detailing the nature of  the boundary conditions.
    }
    \label{fig0}
\end{figure}

Consider  a  2D rectangular  body  $\Omega =[-L,\,L]\times[0,\,H]$. Denote by $\vect{x}\in\Omega$   points  in the reference configuration and by
$\vect{y}(\vect{x})$ their  deformed position, see Fig.  \ref{fig0}.
Working directly with the  deformation gradient  $\tens{F}=\nabla \vect{y}$  we account  for \textit{geometric} nonlinearities. 
In such an   approach not only the principal stretches  $\lambda_{1,2}$ (the square roots of the eigenvalues of $\tens{F}^T\tens{F}$),  can be large   but also that the description of  rotations  is   geometrically exact \cite{OgdenBook}.

 Assume that the  material  is incompressible, so that $\det\tens{F} =\lambda_1\lambda_2=1$, and isotropic, so that the elastic energy density can be written as $\hat w(\lambda_1)=w(\lambda_1,\lambda_1^{-1})$.
We can then write the force balance  in the form
$\nabla\cdot\tens{P}=0,$
where $ P_{ij} $ are the components of the first Piola-Kirchhoff stress tensor
$\tens{P}=  \partial w/\partial \tens{F}+p \tens{F}^{-1}  $
and   $p$   is  the Lagrange multiplier enforcing the incompressibility constraint.

Suppose further that the body $\Omega$ is loaded in a two-sided hard device, such that $y_1= \lambda x_1$ at   $ x_1= \pm L$,  where $\lambda $ is the applied stretch which serves as the control parameter.
Then on the  side boundaries (at   $x_1= \pm L$)   the horizontal displacements are prescribed $y_1 = \pm \lambda L$ while  the possibility of free sliding  is ensured by the second condition
$ P_{12} = {0} $. The upper boundary  $x_2=0$  will  be kept free  so that  $P_{22}=P_{21} = {0}$ while   the lower boundary  $x_2 =H$ will be constrained only partially  so that  $y_2= H/\lambda$  and $P_{21} = 0$.
The ensuing basic   problem  of elasticity theory admits a homogeneous  solution $ \bf{y}^{(0)}\coloneqq\tens{F}^{(0)}\bf x$, where $\tens{F}^{(0)} = \diag{(\lambda,\lambda^{-1})}$; the corresponding pressure is $  p^{(0)} \coloneqq  - \lambda^{-1} \partial w/\partial \lambda_2$.

To study the stability of this solution, we use standard methods \cite{HH75,cao2012wrinkles,holland2017instabilities,li2023periodic} and write   the  perturbed  displacement and pressure fields, in the form $\vect{y} = \vect{y}^{(0)}+ \sum_{j=1}^\infty \epsilon^j\vect{u}^{(j)}$ and $p = p^{(0)} +\sum_{j=1}^\infty \epsilon^j p^{(j)}$ where $\epsilon$ is a small parameter.
Inserting these expansions in the force balance  we obtain, at the first order,  a linear boundary value problem for $\vect{u}^{(1)} $ and  $p^{(1)}$.  

To illustrate the results we introduce the stream function $\vect{u}^{(1)}(\bf x) = (\partial_2 \chi,\,-\partial_1 \chi)$, and write the solution of the first order equilibrium problem in the form 
$\chi = i A g(\gamma x_2) \exp(i  \gamma  x_1)/\gamma + \text{c.c.}$, where $A$ is  still undefined complex amplitude and $\text{c.c.}$ denotes complex conjugate. Here we have also introduced the horizontal wavenumber  $\gamma = (n \pi)/(2\lambda L)$, where $n$ is an integer with  even (odd) values   representing symmetric (asymmetric) modes, respectively. The  expression for $p^{(1)}(\bf x)$ in terms of $g(\gamma x_2)$  is too long to be presented here, see \footnote{See Supplemental Material [url] for the details of the computations, which
includes Refs. \cite{charru2011hydrodynamic,fu2015buckling,alnaes2015fenics,riccobelli2021rods,balay2016petsc,Chadwick_1971}}.

Following closely \cite{HH75}, we  write  the  real valued function  $g$ in the form $g(\gamma x_2)= \sum_{k=1}^4 C_k {\rm exp}[\gamma \omega_k \ x_2] $, where   $\omega_1= -\omega_2=  \alpha$, $\omega_3= -\omega_4=  \beta$.  The constants $\alpha$, $ \beta$  can be found from  the relations $\alpha \beta=\lambda^2$ and $\alpha^2 + \beta^2 +2= \lambda(\lambda^4-1)\eta$; the elastic energy enters these relations  through the function $\eta(\lambda)=\hat w''(\lambda)/\hat w'(\lambda)$ which characterizes the \textit{physical} nonlinearity.

The bifurcation points $\lambda_n(H/L)$, parametrized by the integers  $n(H/L)$,   can be  found from the condition that  there exists a nontrivial  set of coefficients $C_k$,  such that the  functions  $ \vect{u}^{(1)}$ and $p^{(1)})$  satisfy the  boundary conditions at  the linear order.  This gives    an explicit  nonlinear algebraic equation, see \cite{Note1}.  We can then define $ \lambda_\text{cr}(H/L) =\min_{n\geq 1} \lambda_n(H/L)$ and denote by  $n_\text{cr}(H/L)$  the corresponding critical mode. To illustrate the     sensitivity of the  instability threshold  $ \lambda_\text{cr}(H/L)$  to the geometry of the domain characterized by the ratio $H/L$,  we need to choose a specific  energy density. 

\begin{figure}[h!]
    \centering
    \includegraphics[width=\columnwidth]{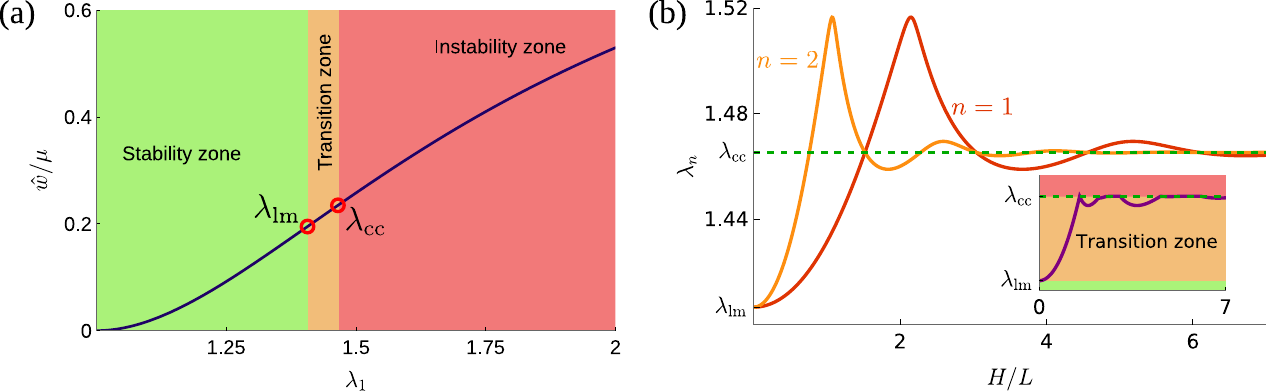}%
    \caption{(a) The energy density $\hat w(\lambda)$ of our softening material as a function of the maximal principal stretch $\lambda_1$.
        (b) The  stability curves for the two modes with  $n=1,\,2$; the purple line in the inset represents the function $\lambda_\text{cr}(H/L)$.
}%
    \label{fig11}
\end{figure}

To account for  strain softening in the simplest form, we assume that 
$ w  = \mu\,(I - 2)/I,$
where  $I = \lambda_1^2 + \lambda_2 ^2$ is the first strain invariant and $\mu$ is the measure of rigidity (see more about this particular choice in \cite{Note1}). In this case $\hat w(\lambda)=\mu(\lambda^2-1)^2/(2(\lambda^4+1))$ and  the  softening ($\hat w'' <0$) takes place for $\lambda > \lambda_\text{lm}=\sqrt[4]{(1/3) \left(\sqrt{33}+6\right)}$, see Fig.~\ref{fig11}(a) and \cite{Note1}. The  value $\lambda_\text{lm}$   is known as the Considère or the load maximum (LM)  threshold  \cite{yasnikov2014revisiting, fielding2011criterion, li2023periodic},
where by the  `load' we understand  the axial stress  in the direction of traction $P(\lambda) = \vect{e}_1\cdot\tens{P}\cdot\vect{e}_1=\hat w'(\lambda)$; reaching this threshold indicates the occurrence of necking in slender bodies \cite{HH75,tvergaard1993necking, audoly2016analysis, needleman2018effect};  it can  be shown that crossing the LM threshold is also a necessary condition for the occurrence of a  general instability  \cite{HH75}. 

Observe first that, independently of the value of $n$, the  functions  $\lambda_n(H/L)$,   shown  in  Fig.~\ref{fig11}(b) for $n=1,2$,   approach the point  $\lambda_\text{lm}\simeq 1.407$ in the limit   infinitely small aspect ratios ($H\ll L$,  thin domains)  
and the point $ \lambda_\text{cc}\simeq 1.465$  in the limit of infinitely large aspect ratios ($H\gg L$, thick domains).

Note that the  emerging threshold $\lambda_\text{cc}$ indicates the failure of the  complementing condition (CC) \cite{ShilavyBook, liu2018boundary, simpson2020complementing, negron2012violation,biot1963surface}. In an infinite system it marks  the onset of wrinkling instability with  all wave numbers becoming unstable simultaneously.  In our case  the value of the CC threshold can be found analytically as a solution of the transcendental equation   $\eta(\lambda_\text{cc})= -\lambda_\text{cc}^{-3}$ \cite{Note1}. Note also that in the classical geometrically  linearized elasticity theory, where  both stretches  and rotations are   small  and  therefore we  can use the approximation
$w(\tens{E})$ with  $\tens{E}= (1/2) (\nabla\vect{u}+\nabla\vect{u}^T)$, the very difference between the thresholds  $\lambda_\text{cc}$ and  $\lambda_\text{lm}$ disappears  and the whole complexity of the emerging  stability diagram is  lost \cite{Note1}.

\begin{figure}[h!]
    \centering
    \includegraphics[width=0.6\columnwidth]{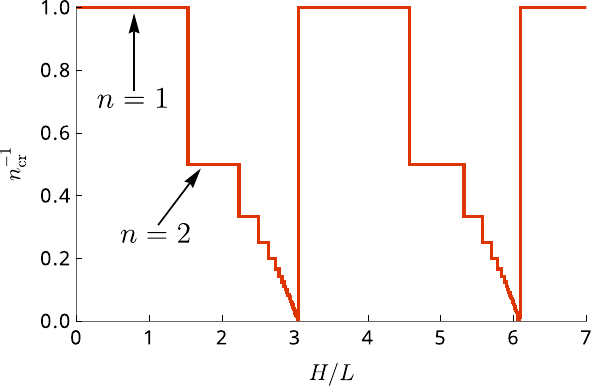}%
    \caption{ The inverse of the critical mode $n_\text{cr}$ versus the aspect ratio $H/L$.The accumulation points correspond to $n= \infty$.}%
    \label{fig12}
\end{figure}

Outside these two   limits (of  infinitely thin and infinitely thick domains),  the behavior of the  function  $\lambda_\text{cr}(H/L)$ looks uncorrelated. However, a remarkable  underlying   structure reveals itself if we focus instead on  the  integer valued function $n_\text{cr}(H/L)$, see  Fig.\ref{fig12}.

First of all, we observe that the  necking-type  instability with $n_\text{cr}=1$ is not a feature of slender bodies only, but appears  periodically as one changes the aspect ratio. 
 Similarly, the  wrinkling-type instability with $n_\text{cr}=\infty$ appears
at  periodically distributed  values of the aspect  ratio. In both cases the period is the same  and is equal to    $ \Delta (H/L) = 4 \lambda^3_\text{cc}/\sqrt{-1+2\lambda _\text{cc}^2+3\lambda_\text{cc}^4}$, see \cite{Note1} for details. %

Overall, we observe a periodic distribution of `staircase' structures with infinite number of steps in every period representing  all integer values  of  $n_\text{cr}$ from necking with $n_\text{cr}=1$ to wrinkling with  $n_\text{cr}=\infty$.  Each of these `staircases' demonstrates  the same  `devilish' features with step  accumulation  taking place around the  recurrent wrinkling thresholds (where  the unstable mode becomes singularly localized near the free surface). In other words, each     `staircase ' describes a  scale-free  crossover between necking and wrinkling with the steps   emerging due to the locking in the parameter intervals  where    horizontal and vertical oscillations of the   displacement field  are resonant with the domain geometry. To the best of our knowledge, the  reported  extreme sensitivity of the critical wave number  to the aspect ratio and the emergence of special  geometries where the  instability pattern changes dramatically from fully  localized   to fully de-localized, have not been previously observed in nonlinear elasticity problems.

The revealed  distribution of the stability thresholds  can be corroborated analytically using  the observation that for $H/L \gg 1$ (when $\lambda_\text{cr} \sim \lambda_\text{cc}$)   one can approximate  the actual   problem of finding $n_\text{cr}(H/L)$,   involving  minimization of  an implicitly given function over a discrete set,  by a model  problem $N = {\mathrm{arg}\,\max}_{\xi}(\sin(a\xi)/e^\xi)$, where  $\xi$ is a positive integer and $a \sim H/L $. The model  problem can be solved explicitly  and its solution   $\color{blue}N$  can be formally proved  to exhibit the periodic staircase structure of the type shown in  Fig.~\ref{fig12}. 

To   determine the nature of detected bifurcations,  we now perform  a standard    weakly nonlinear   amplitude expansion   \cite{becherer2009probing, weinstein1981nonlinear, audoly2011localized, pomeau1983dislocation, koiter1967stability, budiansky1974theory, stuart2nonlinear, vdh2008}.
The idea is to  compute   the next terms of the perturbative expansion $\vect{u}^{(2)} $, $p^{(2)}$ and use the obtained  information  to    determine   $\lambda$ dependence of the amplitude $A$ near the bifurcation point  $\lambda_\text{cr}$. In this respect the  `near necking' ( single-mode instability) and  the `near wrinkling' (multi-mode instability) regimes function differently.

\begin{figure}[h!]
    \centering
    \includegraphics[width=\columnwidth]{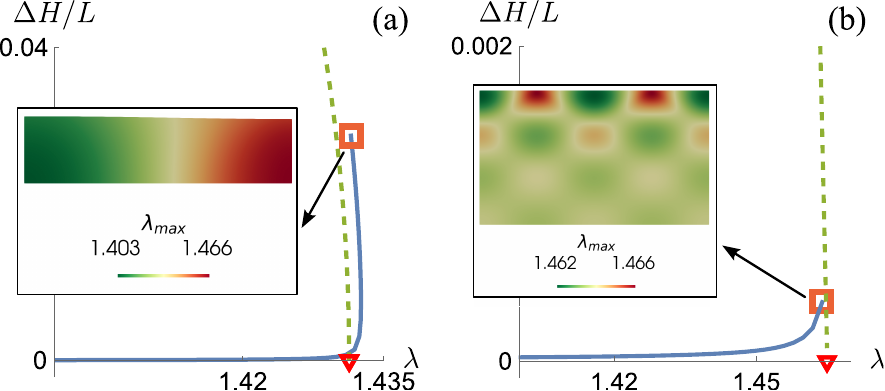}
    \caption{Bifurcation diagrams showing the amplitude $\Delta H$ of the unstable mode on the free surface for the cases: (a)  $H/L=1$    (near necking case) and (b) $H/L=2.5$ (near wrinkling case). The red triangles denote  the critical thresholds $\lambda_\text{cr}$. Solid and  dashed lines represent the  results of the finite element simulations and  of the weakly non-linear analysis, respectively.
        Insets show the distribution of the maximal principal stretch $\lambda_\text{max}$ in the actual configuration corresponding to the location of the square marker.
    }
    \label{fig2}
\end{figure}

Indeed, in the more conventional   `near necking' regimes,  where  the buckling thresholds  $\lambda_n$ are well separated and only a finite number of modes are initially  activated  in the postbuckling regime,  the natural small parameter is known to be $
    \epsilon= \sqrt{|\lambda-\lambda_\text{cr}|/\lambda_\text{cr}}.
$
By expanding the energy functional $\mathcal{W}=\int_\Omega w\,\d\vect{x}$ in $\epsilon$ we obtain \cite{Note1}
$\Delta\mathcal{W}
    = \epsilon^4  ( \theta_{2} \left| A\right|^2 +
    \theta_{4} \left| A\right|^4  ) + o(\epsilon^4)$,  where  $\theta_{2}(\lambda) $, $\theta_{4} (\lambda)$ are known   real functions. The requirement of  stationarity of the energy in $A$ (at order $\epsilon^4$),  gives  the    expression  for the amplitude $A= \sqrt{-\theta_2/(2\theta_4)}$ where  $\theta_2$ and $\theta_4$ have the same sign.  This characterizes the bifurcation as a subcritical (unstable) pitchfork, see the dashed line in Fig.~\ref{fig2} (a).  The implied  unstable postbuckling  regime is  the   diffuse  necking illustrated  in the inset in Fig.~\ref{fig2}(a).

The `near wrinkling'  regimes, where buckling thresholds accumulate,  are  markedly different. In this case
a small increment of the control parameter $\lambda$ away from the critical value $\lambda_\text{cr}$ activates an essentially  infinite number of  instability modes. Therefore  in the weakly non-linear  approximation an unstable mode interacts with many  other  modes. The availability of a broad  bandwidth of  such  modes  requires  a different scaling  and one can show that the   natural small parameter in this case is 
$
    \epsilon= |\lambda-\lambda_\text{cr}|/\lambda_\text{cr},
$ see \cite{fu1999nonlinear, ciarletta2015semi} for similar analyses.
To take into account all the implied  interactions   we  need to modify the   expression    for the first order stream function adopted in the 'near necking' case and write instead  $
    \chi = \sum_{m=-\infty}^{+\infty}i (A_m/\gamma m) g(\gamma m x_2) \exp(i  \gamma m x_1) + \text{c.c.}
$
where  $m$ is an integer and
$A_m$ is amplitude of the mode $m$.
We can then proceed as before  and  find the amplitude equation, accounting for  cubic resonances, which now takes  the form of an infinite system:
$
    \theta_1  A_m +\sum\limits_{k=-\infty}^{+\infty}\theta_{3(k)}A_k A_{m-k}=0.
$
Here the real functions  $\theta_1(\lambda;m)$ and $\theta_{3(k)}(\lambda;m)$ are known  explicitly \cite{Note1}.  The analysis shows that the bifurcation is   again a  subcritical pitchfork, see the dashed line in Fig.~\ref{fig2}(b), which implies that the incipient  postbifurcational mode, illustrated in the inset in Fig.~\ref{fig2}(b) is again unstable.

To complement this  analytical  study  we also performed some parallel direct   numerical simulations.  For numerical convenience  we slightly modified  the model by introducing into our original  energy density $ w(\tens{F})$ a dependence on $J=\lambda_1\lambda_2$, a measure of volumetric deformation.  More specifically we used the expression  $
    w_J(\tens{F})  = (\mu/I)\left(I-2\log J-2\right) + (\Lambda/2)(\log J)^2$  with   $\Lambda$   equal to $100 \mu$  which corresponds to  almost  incompressibility; note that   at  $\Lambda \rightarrow +\infty$ we recover both  the original model.

The bifurcated branch  was recovered after we   introduced a small imperfection on  the  free  boundary  with a wavenumber of  the instability  mode and a small  amplitude  of the order of $10^{-5}\,L$, see the blue lines in Fig.~\ref{fig2}(a,b). We   used an arclength continuation method \cite{seydel2009practical, su2023tunable} which allowed us to reach the state of    strain focusing caused   the local violation of the complementing condition. The   deformation patterns   at  such  limits  (of   the applicability of continuum elasticity) are illustrated in the insets in Fig.~\ref{fig2}(a,b) for the typical `near necking' and 'near wrinkling' regimes.

The ultimate strain localization, which caused the break down of our continuum model, is indicative  of  the trend towards the formation of atomically sharp cracks. To capture the latter, the scale-free continuum theory,  which is  expected to be operative  only on  long waves, can  be regularized through the introduction of a  sub-continuum length scale. A convenient   approach of this type is a  phase-field model of fracture, e.g. \cite{BOURDIN2000797,Karma01,PhysRevLett.85.118,eastgate2002fracture}.  Specifically, we  assume  that %
 $$w_\text{pf}(\tens{F},\alpha) = (1-\alpha)^2(\mu/2)(I-2)+\mu\alpha^2 +\mu\,\ell_0^2\|\nabla \alpha\|^2,$$
   where  $\alpha(\vect{x}) \in[0,1]$ is a subcontinuum  damage-like scalar field: the compatibility with   our original nonlinear elasticity model is ensured by the fact that $w(\tens{F}) =\min_{\alpha\in[0,1]}[(1-\alpha)^2(\mu/2)(I-2)+\mu\alpha^2]$. The regularization is achieved  through the term  penalizing    gradients of   $\alpha$ which brings  an  internal length scale  $\ell_0$. At $\ell_0\ll L$ this approach  is known to be equivalent  to  the  Griffith fracture model with the  toughness $G_c=\mu\ell_0/2$ \cite{griffithphenomena,BOURDIN2000797,tanne2018crack}.

\begin{figure}[h!]
    \centering
    \includegraphics[width=0.65\columnwidth]{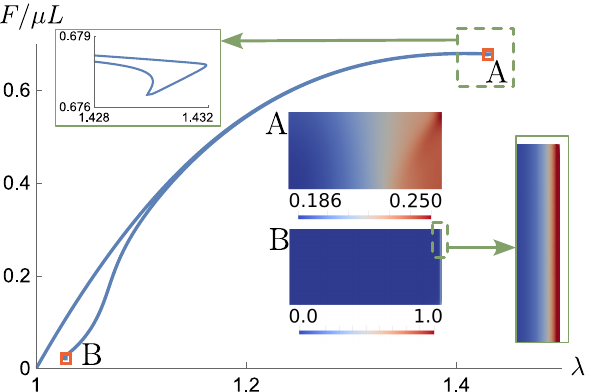}
    \caption{Normalized axial force $F/\mu L$ versus the mean stretch $\lambda$ for the near necking case ($H/L=1$).
        The insets on the right show the distribution of the  internal  variable $\alpha$ in the reference configuration corresponding to the points A and B. The parameter $\ell_0/H=0.01$.
    }
    \label{fig3}
\end{figure}
Going in this way beyond continuum elasticity and adopting again the weak compressibility regularization, we performed a series of numerical simulations  with the goal to capture  the  actual formation of cracks.  We used a Newton's algorithm complemented by a standard pseudo-arclength continuation technique \cite{seydel2009practical} to minimize  at each value of the loading parameter $\lambda$  the energy with respect to both, the deformation  field  $\vect{y}(\vect{x})$ and the auxiliary scalar field   $\alpha (\vect{x})$.  

\begin{figure}[h!]
    \centering
    \includegraphics[width=0.65\columnwidth]{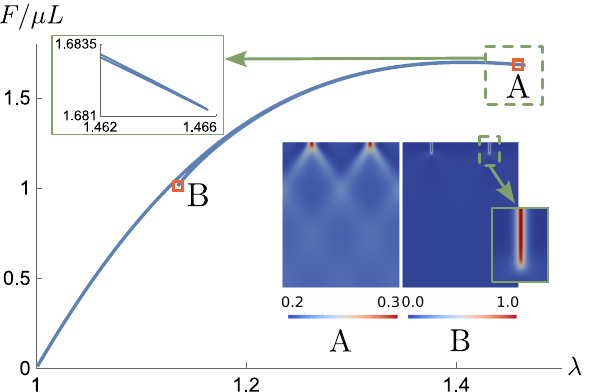}
    \caption{Normalized axial force $F/\mu L$ versus the mean stretch $\lambda$ for the near wrinkling case ($H/L=2.5$).
        The insets on the right show the distribution of the internal variable $\alpha$ in the reference configuration corresponding to the points A and B. The parameter $\ell_0/H=0.01$.
    }
    \label{fig4}
\end{figure}

The results of the two representative numerical simulations, illustrating qualitatively different `near necking' and `near wrinkling' regimes,  are presented in Figs.~\ref{fig3}-\ref{fig4}.
   In both figures the (unstable) post-bifurcational response is  represented  through the dimensionless  force-stretch relation
$F (\lambda) = \int_{-H}^{0}\left. P_{11}(\lambda)\right|_{x_1=L}\,\d x_2.$ The deformed configurations  close and far  from the bifurcation points are shown  in the insets.
  Note that while in our `near wrinkling' regime we show for simplicity only the case with  two emerging cracks,  the aspect ratio of the domain and the regularization length could be chosen differently to obtain arbitrary many cracks.

The common feature of the two cases,  shown in Figs.~\ref{fig3}-\ref{fig4}, is  the  gradual  sharpening  of the initially diffuse local  'non-affinity' measured by parameter $\alpha$.   The actual formation of cracks   can be linked to the moment of reaching the value $\alpha \sim 1$  inside  the  localized  regions with  the  thickness   of  order of sub-continuum scale  $\ell_0$. Since the focus of our study is    crack nucleation,  we did not advance  our simulations   till the   complete break down of the slab which is preceded  by secondary bifurcations representing both crack branching and crack arrest \cite{bourdin2014morphogenesis}.  Overall, the presence in this problem of a subcritical bifurcation indicates the possibility of abrupt (dynamic) transition from a homogeneous state to a cracked state which is a  typical scenario in brittle fracture. 

To conclude, using the  simplest geometrical setting and focusing on initially flawless soft solids,  we   showed that crack nucleation is  preceded by an elastic instability which can be identified  using  continuum elasticity theory  only if the latter  accounts properly for both geometrical  and physical  nonlinearities. Such elasticity  theory predicts  a surprisingly complex   linear stability  diagram with recurrent  geometry-sensitive  crossovers between necking and wrinkling modes. Both necking and wrinkling instabilities were shown to   evolve  towards the formation of developed cracks when the classical elasticity was seamlessly  extended   as  a phase-field type  model. Our analysis builds  a   bridge between  nonlinear elasticity and fracture mechanics and points to  the existence of  purely elastic precursors of crack nucleation. Similar mechanisms should be operative in other highly nonlinear manifestations of elasticity such as   cavitation \cite{ball1982discontinuous, kumar2021poker}, phase nucleation \cite{healey2007two,grabovsky2013marginal}, and creasing \cite{ciarletta2019soft,pandurangi2022nucleation}. 

\acknowledgments{DR and PC gratefully acknowledge partial support from grant Dipartimento di Eccellenza 2023-2027 and MIUR, PRIN 2017 Project ``Mathematics of active materials: From mechanobiology to smart devices''. DR has been partially supported by MIUR, PRIN 2022 Project ``Mathematical models for viscoelastic biological matter''. DR and PC are members of the INdAM research group GNFM. This work was performed using HPC resources from GENCI-IDRIS (Grant 2023-AD010913451R1).}

\end{document}


\title{Elastic instability behind  brittle fracture: Supplemental Material}

\author{D. Riccobelli}
\affiliation{MOX -- Dipartimento di Matematica, Politecnico di Milano, 20133 Milano, Italy}
\author{P. Ciarletta}
\affiliation{MOX -- Dipartimento di Matematica, Politecnico di Milano, 20133 Milano, Italy}
\author{G. Vitale}
\affiliation{Laboratoire de Mecanique des Solides, École Polytechnique, 91128 Palaiseau, France.}
\author{C. Maurini}
\affiliation{CNRS, Institut Jean Le Rond d’Alembert, Sorbonne Université, UMR 7190, 75005 Paris, France.}
\author{L. Truskinovsky}
\email{lev.truskinovsky@espci.fr}
\affiliation{ESPCI ParisTech, PMMH, CNRS -- UMR 7636, 75005 Paris, France.}

\graphicspath{{figures}}
\maketitle
\onecolumngrid

\subsection{Linear stability analysis}  At the linear order  the incremental Piola--Kirchhoff stress is of the form 
$
\tens{P}^{(1)} = \mathbf{K}^{(1)}:\tens{\Gamma} - p^{(0)} \tens{\Gamma} + p^{(1)} \tens{I}
$
where $\tens{\Gamma} = \nabla_\vect{x}\vect{u}^{(1)}$, $\nabla_\vect{x}$ is the gradient operator in the finitely deformed configuration $[-\lambda L,\,\lambda L]\times[0,\,H/\lambda]$, and $\mathbf{K}^{(1)}$ is the tensor of instantaneous elastic moduli. The components of $\mathbf{K}^{(1)}$ are given by \cite{OgdenBook}
\[
\begin{aligned}
K^{(1)}_{iijj} &= \lambda_i\lambda_j\frac{\partial^2 W}{\partial \lambda_i\partial \lambda_j},\\
K^{(1)}_{ijij} &= \frac{\left(\lambda_i\dfrac{\partial W}{\partial \lambda_i}-\lambda_j\dfrac{\partial W}{\partial \lambda_j}\right)\lambda_{i}^2}{\lambda_i^2-\lambda_j^2}, &&i\neq j,\\
K^{(1)}_{ijji} &= K^{(1)}_{jiij} = K^{(1)}_{ijij}-\lambda_i\frac{\partial W}{\partial \lambda_i}, &&i\neq j,\\
\end{aligned}
\]
for $i,\,j=1,\,2$.
At the linear order, the incremental equations can be written as
$ \nabla_{\vect{x}} \cdot \tens{P}^{(1)} = 0,\,\,\,
 \tr \tens{\Gamma} = 0, $
which are complemented by the linearized form of the boundary conditions:
$\vect{u}^{(1)}\cdot\vect{e}_2 = 0$ at$x_2=H/\lambda$;
$\vect{e}_1\cdot{\tens{P}^{(1)}}^T\vect{e}_2 = 0$ at $x_2=H/\lambda$;
$\vect{u}^{(1)}\cdot\vect{e}_1 = 0$ at $x_1=\pm\lambda L$;
$\vect{e}_2\cdot{\tens{P}^{(1)}}^T\vect{e}_1 = 0$ at  $x_1=\pm\lambda L$ and finally 
$\tens{P}^{(1)}\vect{e}_2=\vect{0}$ at $x_2=0$.
As explained in the main text, it is natural to look for solutions in the form
\begin{equation}
u^{(1)}_1=\imath A g'(\gamma  x_2) e^{\imath  \gamma  x_1 }+ c.c., \qquad u^{(1)}_2= A g(\gamma  x_2) e^{\imath  \gamma  x_1 }+ c.c.,
\label{u1}
\end{equation}
\begin{equation}
p^{(1)}=\gamma A ((K^{(1)}_{1122} + K^{(1)}_{2112} - K^{(1)}_{1111})g'(\gamma x_2) + K^{(1)}_{2121}g'''(\gamma x_2)) e^{\imath  \gamma  x_1 }+ c.c.,
\label{p1}
\end{equation}
with the scalar function $g$   given in this general form 
\begin{equation}
   \label{eq:g}
g(\gamma x_2) = C_1 \cosh (\alpha  \gamma  x_2)+ C_2 \sinh (\alpha  \gamma  x_2)+ C_3 \cosh (\beta  \gamma  x_2)+ C_4 \sinh (\beta  \gamma  x_2).
\end{equation}
Here $C_1,\,C_2,\,C_3,\,C_4$ are constants to be found from  the boundary conditions, while $\alpha$ and $\beta$ satisfy 
\begin{equation}
\label{eq:ab}
	\a^2 = \frac{b}{c} + \sqrt{\left(\frac{b}{c}\right)^2 - \frac{a}{c}},\,\,\,  
   \b^2 = \frac{b}{c} - \sqrt{\left(\frac{b}{c}\right)^2 - \frac{a}{c}},
\end{equation}
where 
\begin{align}
\begin{split}
    a &= K^{(1)}_{1212} = \ds \lambda^5\frac{w'(\lambda)}{\lambda^4 - 1}, \,\,\
	 c = K^{(1)}_{2121} = \ds \lambda\frac{w'(\lambda)}{\lambda^4 - 1}, \\
	2b &= K^{(1)}_{1111} + K^{(1)}_{2222} - 2(K^{(1)}_{1122} + K^{(1)}_{2112})= \ds \lambda^2 w''(\lambda) - 2c.
\end{split}
\end{align}
The non trivial solutions of the resulting linear system exists  whenever the following characteristic equation for the eigenvalue $\gamma$ is satisfied
\begin{equation}
\begin{gathered}
\label{eq:charcteristic}
\left(\alpha ^4 \beta +2 \alpha ^3 \beta ^2+2 \alpha ^2 \left(\beta ^3+\beta \right)+\alpha  \left(\beta ^4+2 \beta ^2-1\right)-\beta \right) (\alpha -\beta )^2 \sinh \left(\frac{H \gamma  (\alpha +\beta )}{\lambda }\right)+\\
+(\alpha +\beta )^2 \left(\alpha ^4 \beta -2 \alpha ^3 \beta ^2+2 \alpha ^2 \left(\beta ^3+\beta \right)-\alpha  \left(\beta ^4+2 \beta ^2-1\right)-\beta \right) \sinh \left(\frac{H \gamma (\alpha-\beta) }{\lambda }\right)=0.
\end{gathered}
\end{equation}
The corresponding  eigenfunction $g$ can be then written in the form:
\begin{equation}\label{eigenfctn}
 \begin{split}
 g(\gamma x_2) = \cosh(\a\gamma x_2) -
 \frac{\cosh(\a\gamma h) - \frac{\a^2 + 1}{\b^2 + 1}\cosh(\b\gamma H/\lambda)}{\sinh(\a \gamma H/\lambda) - \frac{\b^2 + 1}{\a^2 + 1}\frac{\a}{\b}\sinh(\b\gamma H/\lambda)} \cosh(\b\gamma x_2) +\\
 - \frac{\a^2 + 1}{\b^2 + 1}\sinh(\a\gamma x_2) +
 \frac{\frac{\b^2 + 1}{\a^2 + 1}\cosh(\a\gamma H/\lambda) - \cosh(\b\gamma H/\lambda)}{\frac{\a}{\b}\sinh(\a \gamma H/\lambda) - \frac{\b^2 + 1}{\a^2 + 1}\sinh(\b\gamma H/\lambda)}\sinh(\b\gamma x_2).
 \end{split}
 \end{equation}

\subsection{ Material behavior} 
To achieve analytical transparency, we consider the strain energy  density of a softening solid in the following simple form
\begin{equation}
   \label{eq:elastic_strain_energy}
w(\tens{F}) = \frac{\mu}{I}(I-2) = \mu \frac{\left(\lambda_1 ^2-1\right)^2 }{\lambda_1 ^4+1},
\end{equation}
where $I = \tens{F}:\tens{F} = \lambda_1^2+\lambda_1^{-2}$.
For the homogeneous deformation $\tens{F}^{(0)}=\diag(\lambda,\,\lambda^{-1})$, the strain energy density reduces to the function $w(\lambda)$. The  load maximum (LM) stretch  $\lambda_\text{lm}$ introduced in the main text  and corresponding to the maximum of $w'(\lambda)$ is now $\lambda_\text{lm}=\sqrt[4]{\frac{1}{3} \left(\sqrt{33}+6\right)}$.
The complementing condition (CC) is violated whenever the incremental problem admits a non-trivial solution  in the half-space. Therefore, by taking the limit $H\rightarrow+\infty$, 
in the characteristic equation  \eqref{eq:charcteristic} 
we obtain an equation for $\lambda=\lambda_\text{cc}$:
\begin{equation}
   \label{eq:CC}
\lambda_\text{cc} ^3 w''(\lambda_\text{cc})+w'(\lambda_\text{cc} )=0.
\end{equation}
In   the case of the strain energy density \eqref{eq:elastic_strain_energy}, the  equation \eqref{eq:CC} reads
$
3 \lambda_\text{cc} ^{11}-\lambda_\text{cc} ^9-12 \lambda_\text{cc} ^7+\lambda_\text{cc} ^3+\lambda_\text{cc}=0,
$
and  its relevant root is $\lambda_\text{cc}\simeq1.46527$.

The formal advantage of the choice  \eqref{eq:elastic_strain_energy} for   $W(\tens{F})$ becomes  clear at the very end of our study where we take advantage of the fact that such an energy density  can be obtained as the minimum of a function  involving the auxiliary field $\alpha$ characterizing material softening:
\begin{equation}
   \label{eq:psi}
w(\tens{F}) = \min_{\alpha\in [0,1]} \psi(\alpha,\,\tens{F}) =  \min_{\alpha\in [0,1]} a(\alpha)\widehat{w}(\tens{F}) + \phi(\alpha),
\end{equation}
where $ \widehat{w}(\tens{F}) = \frac{\mu}{2}(I - 2) = \frac{\mu}{2}\left(\lambda_1^2+\lambda_1^{-2}-2\right)$  is the strain energy density of a standard neo-Hookean material,  $a(\alpha) = (1-\alpha)^2$ is the measure of stiffness degradation and    $\phi(\alpha)=\mu \alpha^2$ is the energetic  price of such degradation.

\subsection{Periodicity of  stability thresholds}
Assume that the thresholds $\lambda_\text{cc}>\lambda_\text{lm}>1$ exist  for a given strain energy density $w(\lambda_1)$ 
 and suppose  that  $\lambda\in[\lambda_\text{lm},\,\lambda_\text{cc}]$. Then   from \eqref{eq:CC} we have 
$
\eta(\lambda)\in\left[- 1/\lambda ^3 ,\,0\right].
$
Then $(\alpha+\beta)^2>0$, $(\alpha-\beta)^2<0$, while $\alpha\beta$ is real and positive. Therefore,   $\alpha$ and $\beta$ are complex conjugate numbers and we can write
$
\alpha = \gamma +\imath\delta,\qquad\beta = \gamma-\imath\delta,\qquad\gamma,\,\delta>0,
$
with $\delta = \delta(\lambda)$, $\gamma=\gamma(\lambda)$, whose  explicit expressions can be easily obtained.
The characteristic equation \eqref{eq:charcteristic} can be now rewritten as
\begin{equation}
    \label{eq:charact_CC}
\sin(\tilde{A}n) = f(\lambda,\,n).
\end{equation}
where
\begin{equation}
    \label{eq:Af}
\tilde A(\lambda) = \frac{\pi  \delta(\lambda) }{ \lambda ^2}\frac{H}{L},\qquad
f(\lambda,\,n)=\frac{\delta  \left(1+\lambda ^3 \eta (\lambda )\right) }{\gamma  \left(1-\lambda ^3 \eta (\lambda )\right)}\sinh \left(\frac{\pi  \gamma   n}{ \lambda ^2} \frac{H}{L}\right).
\end{equation}
We observe that $f(\lambda_\text{cc},\,n)=0$ for any value of $n$. The corresponding values of the aspect ratio are
\[
\frac{H}{L} = \frac{m \lambda^2}{\delta(\lambda_\text{cc})} = \frac{2 m \lambda^3_\text{cc}}{\sqrt{3\lambda_\text{cc}^4+2\lambda _\text{cc}^2-1}}.\qedhere
\]

\subsection{Auxiliary model}
The equation  \eqref{eq:charact_CC} is not transparent but it can be dramatically simplified for $\lambda \approx \lambda_\text{cc}$. Indeed, 
let $F(\lambda) = \sin(An)-f(\lambda,\,n)$, and rewrite  \eqref{eq:charact_CC} in the form
$
F(\lambda)= 0.
$
By expanding this equation  around  $\lambda=\lambda_\text{cc}$ we obtain
$
F(\lambda_\text{cc},n)+F'(\lambda_\text{cc},n)(\lambda-\lambda_\text{cc})=0 
$
from where   the critical wavenumber can be computed as
\begin{equation}
    \label{eq:ncrasymp}
n_\text{cr} =\argmax_{\mathbb{N}^+}\left(\frac{\gamma (\lambda_\text{cc})^2 \sin (n \tilde A(\lambda_\text{cc}))}{n \gamma (\lambda_\text{cc})^2 \tilde A'(\lambda_\text{cc}) \cos (n \tilde A(\lambda_\text{cc}))-\gamma (\lambda_\text{cc}) \delta (\lambda_\text{cc}) S'(\lambda_\text{cc}) \sinh (n B(\lambda_\text{cc}))}\right)
\end{equation}
where
\[
S(\lambda)=\frac{\left(1+\lambda ^3 \eta (\lambda )\right) }{\left(1-\lambda ^3 \eta (\lambda )\right)},\qquad B(\lambda)=\frac{\pi  \gamma (\lambda)}{ \lambda ^2} \frac{H}{L}.
\]
Consider next the limit   $H/L \to \infty$. Then we  can neglect the cosine carrying term in the denominator  of  \eqref{eq:ncrasymp} and approximate the remaining part of the denominator  by the  the exponential term of the form $e^{Bn}$, where $B$ turns our to be an irrelevant parameter as far as the limiting behavior at large $H/L$ is concerned,  with the right asymptotic behavior already recoverable at $B=1$.
\begin{figure}[t!]
    \centering
    \includegraphics[width=0.5\textwidth]{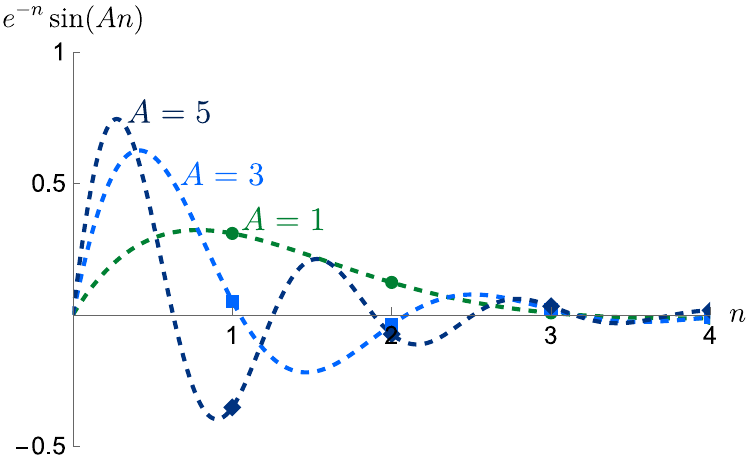}%
    \caption{The function $e^{-n}\sin (A n)$  for $A=1,\,3,\,5$. The markers indicate the values of the functions evaluated for $n\in\Np$.}
    \label{fig:B1}
\end{figure}

We can then  introduce an  auxiliary problem:
\begin{equation} \label{aux}
 N(A)\coloneqq \argmax_{n\in\Np} \left(e^{-n}\sin (A n) \right),
\end{equation}
see Fig.  \ref{fig:B1} showing  the optimized function at several values of parameter $A$. It also illustrates the complexity of the problem \eqref{aux} which implies  integer valued optimization. 
\begin{figure} [ht!]
    \centering
    \includegraphics[width=0.7\textwidth]{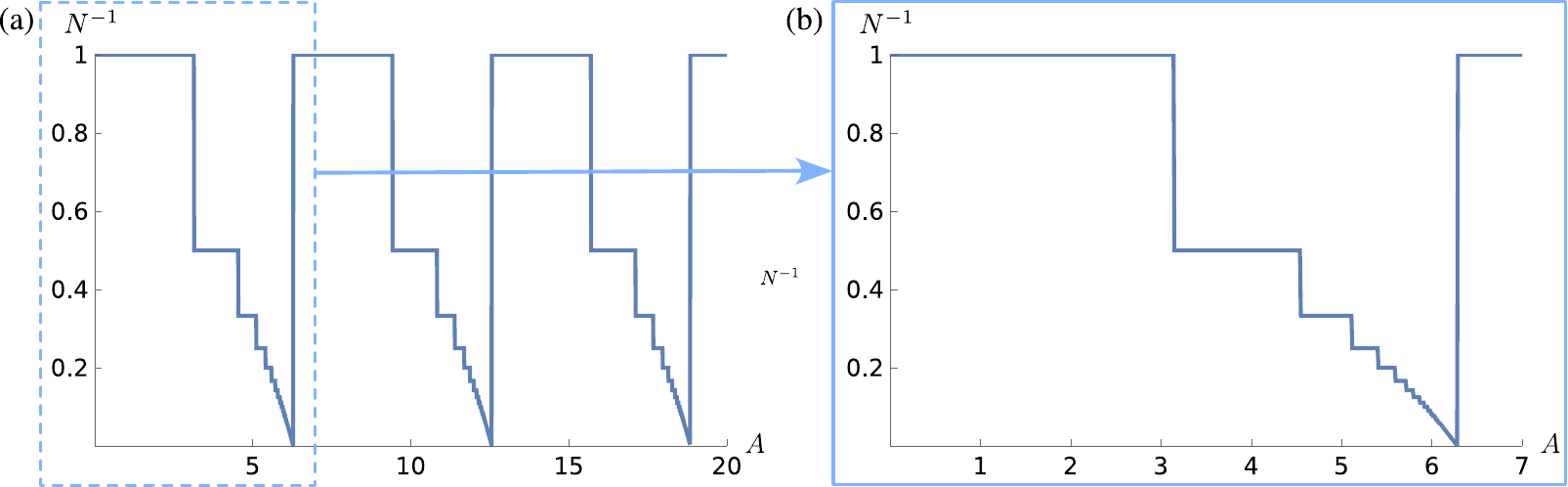}%
    \caption{Plot of the inverse of $N(A)$ over three periods (a) and a zoom over one period (b). 
    }
    \label{fig:B}
\end{figure}

From the boundeness and the periodicity of $\sin(An)$, it follows that the integer valued function $N(A)$ is well defined and non-negative. Since $N(A)$ is periodic with period $2\pi$, it is sufficient to analyze its behavior on the interval $[0,\,2\pi]$.

We first show that $ N(A)\equiv1$ if $A\in [0,\,\pi]$. Indeed,   by induction   in this range of $A$   we have $\frac{\sin(A n)}{\sin A}\leq n$. This is clearly true for $n=1$. 
   Then, if $\sin(An)\leq n \sin(A)$ we obtain
$
    \frac{\sin(A(n+1))}{\sin A} = \frac{\sin (An)\cos A}{\sin A}+\cos(An) \leq n+1.
$
    Thus,  
$
e^{-n}\sin{An}\leq e^{-n}n\sin A\leq e^{-1}\sin A
$
    which proves the result. We next show that  $
    \lim_{A\rightarrow 2\pi^{-}} N(A)  = +\infty.
$  Indeed, for any $M\in\Np$ we can take $\delta=\pi/M$, and then $\sin(An)<0$ for all $n\leq M$ and $A\in (2\pi-\delta,\,2\pi)$. Since the maximum over $n$ of $e^{-n}\sin(An)$ exists and is positive, $ N(A)>M$ for $A\in(2\pi-\delta,\,2\pi)$. This result also proves  the periodic blow up of $ N(A)$ at $A\rightarrow 2m\pi^{-}$ with $m\in\mathbb{N}^+$, which is in agreement with the behavior of the function  $n_\text{cr}$ obtained numerically,  see Fig.~\ref{fig:B}.
Next we show   that the integer valued function $N(A)$, apparently exhibiting `devilish' features around the points $2m\pi^{-}$  in Fig.~\ref{fig:B}, indeed  takes all  positive integer values as the parameter $A$ changes inside the interval of  periodicity,  $A\in[0,\,2\pi]$.  Suppose  that  that $N(A)=m$ at some   $A=A_m$.
If $A_m\in I_m=[2\pi - \pi/m,\,2\pi]$, then $\sin(A_ml)\leq0$ whenever  $l\leq m$ and  $l\in\Np$. Therefore, $N(A)>m$ in $I_m$. Our conjecture follows from the observation that  $N(A)=m+1$ for $A_{m+1}=2\pi-\pi/m$, which can be checked by direct substitution
\[
N (A_{m+1})= \argmax_{n\in\Np} \left(e^{-n}\sin \left(\left(2\pi-\frac{\pi}{m}\right) n\right) \right)=m + \argmax_{l\in\Np} \left(e^{-l}\sin \left( \frac{\pi}{m}l\right) \right) =m+1,
\]
where   we used the fact that $\argmax_{l\in\Np} \left(e^{-l}\sin \left( l\pi/m\right) \right)=1$ since  $\pi/m\in[0,\,\pi]$.

 \subsection{Weakly nonlinear stability analysis}

\emph{ Near necking. }  In this case, the critical threshold $\lambda_\text{cr}$ is a non-degenerate stationary point of the marginal stability curve ($\lambda$,$n$). Accordingly,  a small increase of order  $\epsilon^2$  beyond this  critical threshold actives a narrow bandwidth of order $\epsilon$ of marginally stable modes.  
 Thus, we can use the order parameter $\epsilon= \sqrt{(\lambda-\lambda_\text{cr})/ \lambda_\text{cr}}\ll 1$ to expand  the weakly nonlinear terms in the energy functional. 
The second order equilibrium equations and the incompressibility constraint  take  the form:
\begin{equation}
K^{(1)}_{jikl}u^{(2)}_{k,lj}+K^{(2)}_{jiklnm}u^{(1)}_{m,n}u^{(1)}_{k,lj}+p^{(2)}_{,i} -p^{(1)}_j u^{(1)}_{j,i}=0
\label{gov2}
\end{equation}
\begin{equation}
\diver {\bf u}^{(2)}=  u^{(1)}_{1,2}u^{(1)}_{2,1}-u^{(1)}_{2,2}u^{(1)}_{1,1}.
\label{inc2}
\end{equation}
Here $K^{(2)}_{ijklnm}=\sum_{a,\,b,\,c=1}^2F_{i\alpha}F_{kb}F_{nc}\frac{\partial^3W}{\partial F_{ja}\partial F_{lb}\partial F_{mc}}$ are the third-order instantaneous elastic moduli.
 Similarly, we can obtain the second-order boundary conditions.

The solution of the resulting boundary value problem,   implying   quadratic
 coupling with the first order  solution,   can be written in the form 
$
u^{(2)}_i=U^{(2)}_{i}(\gamma x_2)|A|^2+\left(V^{(2)}_{i}(\gamma x_2)A^2\
e^{2\imath k x_1}+c.c.\right),$
where we introduced the scalar fields  $U^{(2)}_{i}=U^{(2)}_{i}(\gamma x_2)$ and $V^{(2)}_{i}=V^{(2)}_{i}(\gamma x_2)$ which are known  explicitly but the corresponding   expressions are too  cumbersome to be presented    here.

A further series development of the solution would introduce resonant terms in the
third-order problem due to  the cubic  
interactions of the linear modes. The corresponding solutions can be obtained  by requiring that the energy functional is stationary at each relevant order in $\epsilon$   as described it is described in detail for instance in \cite{budiansky1974theory}.

If we now expand  the elastic energy about the homogeneous solution  
\begin{equation}
G(\vect{u})=\mathcal{W}(\vect{u})-\mathcal{W}(\vect{u}^{(0)})=\int_\Omega \left[ p u_{i,i}+\frac{1}{2}K^{(1)}_{jilk}u_{k,l}u_{i,j} +\frac{1}{6}K^{(2)}_{jilknm}u_{k,l}u_{i,j}u_{m,n}+\frac{1}{24}K^{(3)}_{jilknmqs}u_{k,l}u_{i,j}u_{m,n}u_{s,q}\right]\,\d V,
\label{G}
\end{equation}
where  $K^{(3)}_{jilknmqp}$ are the fourth-order instantaneous elastic moduli, and substitute into \eqref{G} the series expansion of the displacement and pressure fields while  using \eqref{gov2}  as well as the incompressibility condition at order $\epsilon^4$ in the form
$
u^{(4)}_{i,i}= u^{(1)}_{j,k}u^{(3)}_{k,j}+\frac{1}{2}u^{(2)}_{j,k}u^{(2)}_{k,j}-\frac{1}{2}(u^{(2)}_{i,i})^2,
$
we obtain :
\begin{equation}
\begin{aligned}
G(\vect{u})=\epsilon^4&\int_\Omega
\left[\frac{1}{2}p^{(0)}u^{(2)}_{j,k}u^{(2)}_{k,j}+\frac{1}{2}p_1 u^{(1)}_{j,k}u^{(1)}_{k,j}-\frac{1}{8}p^{(0)}(u^{(1)}_{j,k}u^{(1)}_{k,j})^2\right.\\
&+ p_1 u^{(1)}_{j,k}u^{(2)}_{k,j} +\frac{1}{2}K^{(1)}_{jilk}u^{(2)}_{k,l}u^{(2)}_{i,j}+\frac{1}{2}K^\star_{jilk}u^{(1)}_{k,l}u^{(1)}_{i,j} +\\
&\left.\frac{1}{2}K^{(2)}_{jilknm}u^{(1)}_{k,l}u^{(1)}_{i,j}u^{(2)}_{m,n}+ \frac{1}{24}K^{(3)}_{jilknmqs}u^{(1)}_{k,l}u^{(1)}_{i,j}u^{(1)}_{m,n}u^{(1)}_{s,q}\right]\,\d V +O(\epsilon^4)
\end{aligned}
\label{G4}
\end{equation}
where $p_1=(dp^{(0)}/d\lambda)|_{\lambda=\lambda_\text{cr}}$,  $K^\star_{jilk}=(\partial K^{(1)}_{jilk} /\partial \lambda)|_{\lambda=\lambda_\text{cr}}$ .
Observe  that the terms of order $\epsilon^3$ vanish due to periodic nature  of the boundary conditions in the horizontal direction.  Note also that the next-to-leading term of order $\epsilon^4$ accounts for the sub-harmonic resonance of the critical mode. In view of the  the periodicity of this mode, it is sufficient to perform the integration in \eqref{G4}   over a critical wavelength. We finally obtain
\begin{equation}
G/\epsilon^4 = \theta_{2} \left| A\right|^2 +
\theta_{4} \left| A\right|^4  +O(\epsilon)
\label{dUnormal}
\end{equation}
where  $\theta_{2}$, $\theta_{4}$ are known real constants.   In particular,  $\theta_{2}$ is a function only of the critical incremental solution, while $\theta_{4}$ also depends on the subharmonic near-critical mode.

The amplitude equation   now follows from the  condition that the incremental  energy $G$ is stationary (at the lowest nontrivial order $\epsilon^4$)  which means that  $dG/dA=0$ and 
$
{ \theta_{2}}  A +{ 2\theta_{4}} A\left| A\right|^2=0.
$
Combined with  its  complex conjugate, this equation shows that the bifurcation is  of a pitchfork type \cite{charru2011hydrodynamic}.  Since  $\theta_2$ and $\theta_4$ are found having  the same sign, the pitchfork is subcritical.

\emph{ Near wrinkling.} 
In this case, the critical threshold $\lambda_\text{cr}$ is a degenerate stationary point of the marginal stability curve ($\lambda$,$n$). Accordingly,  a small increase of order  $\epsilon$  beyond this  critical threshold actives an infinite bandwidth of marginally unstable modes. Thus, the weakly nonlinear analysis requires a different scaling comparing to the  'near necking' case.  By choosing  $\epsilon= \frac{\lambda-\lambda_\text{cr}}{\lambda_\text{cr}}\ll 1$  we  can capture   the 
the superposition of all subharmonic modes representing incremental fields, see  \cite{ciarletta2015semi} for a similar analysis and additional references. Thus, the first order terms have now  the following structure
\begin{equation}
u^{(1)}_1=\imath \sum\limits_{m=-\infty}^{+\infty} A_m g'(\gamma_\text{cr} m x_2) e^{\imath  \gamma m x_1 } , \qquad u^{(1)}_2= \sum\limits_{m=-\infty}^{+\infty} A_m g(\gamma_\text{cr} m  x_2) e^{\imath  \gamma m x_1 },
\label{u1nd}
\end{equation}
\begin{equation}
p^{(1)}=\sum\limits_{m=-\infty}^{+\infty} A_m m \gamma  ((K^{(1)}_{1122} + K^{(1)}_{2112} - K^{(1)}_{1111})g'(\gamma_\text{cr} m x_2) + K^{(1)}_{2121}g'''(\gamma_\text{cr} m x_2)) e^{\imath  \gamma m x_1 },
\label{p1nd}
\end{equation}
where $m$ is an integer and   
$A_m$ is the amplitude of  mode $m$.%

From the linear stability analysis
we know  that  
$A_{-m}=\bar{A}_m$, $A_{0}=0$ and $g(-\gamma_\text{cr} m x_2)=\bar{g}(\gamma_\text{cr} m x_2)$.  By expanding    the  energy functional \eqref{G}, at the next-to-leading order we find
\begin{equation}
G=\frac{\epsilon^3}{2} \int_\Omega\left[ p_1 u^{(1)}_{i,j}u^{(1)}_{j,i}+K^\star_{jikl} u^{(1)}_{i,j}u^{(1)}_{l,k}+p^{(1)}u^{(1)}_{i,j}u^{(1)}_{j,1}+\frac{1}{3}K^{(2)}_{jiklqp} u^{(1)}_{i,j}u^{(1)}_{l,k}u^{(1)}_{p,q}\right]\,\d V + o(\epsilon^3).
\label{Gndsimb}
\end{equation}
where  we account for the fact that cubic resonances between  each of the   infinitely many  linearly unstable modes become dominant at order $\epsilon^3$.  Using   the  solution of the incremental  (first-order) equilibrium problem which we do not present in full detail here  and which specifies the unknown functions in (\ref{u1nd},\ref{p1nd}), we obtain
\begin{equation}
\frac{2G}{\epsilon^3}= \sum\limits_{m=-\infty}^{+\infty}\theta_1(m) |A_m|^2 +\sum\limits_{r=-\infty}^{+\infty}\theta_3(r,m)\bar{A}_mA_r A_{m-r}+ o(1),
\label{Gnd}
\end{equation}
Here $\theta_1(m)$ and $\theta_3(r,\,m)$ are known real constants which are known functionals of the incremental first-order solutions.   We remark that in our case the subharmonic resonance of the critical modes turn out to be  appearing only at order $\epsilon^4$, and can be therefore neglected in the analysis of the amplitude equation.  The latter can be obtained from \eqref{Gnd}  using the same reasoning  as in the 'near necking' case.  Specifically, the stationarity of the energy functional  \eqref{Gnd}   with respect to each amplitude $A_m$ gives  
\begin{equation}
\theta_1(m) A_m +\sum\limits_{r=-\infty}^{+\infty}\theta_3(r,m)A_r A_{m-r}=0,
\label{GLnd}
\end{equation}
which is exactly the amplitude equation presented  in the main text.  To construct the graphs presented in the main paper we solved a truncated system up to a finite order $M$ until a given convergence is reached, adapting the numerical procedure proposed in \cite{fu2015buckling}.

\subsection{Regularized model} 
In our numerical simulations we use  the energy density  accounting for both, phase field type regularization and weak compressibility:
\begin{equation}\label{eigenfctn1}
 \begin{split}
  w_\text{pf}(\alpha,\,\tens{F})  =  a(\alpha)w_J(\tens{F}) + \phi(\alpha) + \phi(1)\ell^2\|\nabla\alpha\|^2, \\
  w_J(\tens{F})  = \frac{\mu}{I}\left(I-2-2\log J\right) + \frac{\Lambda}{2}(\log J)^2.
\end{split}
\end{equation}
We remark that, as the auxiliary field $\alpha$ increases, the shear modulus of the material diminishes but its compressibility is not affected. 
In all the simulations, the parameter $\Lambda$ was set equal to $100 \mu$.

The  fully non linear problem was approximated using the finite element method. We discretize the displacement field by using continuous piecewise quadratic functions  to avoid locking phenomena. On the other hand, the auxiliary field $\alpha$ is approximated by piecewise linear functions.

We decompose the displacement field into two components 
$
\vect{u}(\vect{x}) = \vect{u}^{(0)}(\vect{x})+\delta\vect{u}(\vect{x}),
$
where $\vect{u}^{(0)}$ is the homogeneous solution. This choice is motivated by the fact that we  can impose homogeneous boundary conditions on the displacement field $\delta\vect{u}$, simplifying the implementation of the continuation method.
The deformation gradient  becomes
$
\tens{F} = \tens{I} + \nabla \vect{u} = \tens{I} + \nabla \vect{u}^{(0)}+ \nabla \delta\vect{u}.
$
In order to select  the bifurcated branch in the post-bifurcation regime, a small imperfection is imposed at the boundary of the domain with the amplitude   of the order of $10^{-5}\,L$. The wavenumber of the imperfection is chosen to match the nontrivial eigenfunction of  linear stability analysis.

Then, we use the  arclength continuation method \cite{seydel2009practical} to construct the bifurcation diagram and to study the post-bifurcational behavior.

The method was implemented in Python using the software collection for the numerical solution of partial differential equations FEniCS \cite{alnaes2015fenics}, the library BiFEniCS \cite{riccobelli2021rods}\footnote{\url{https://github.com/riccobelli/bifenics}} for the arclength continuation method, while  PETSc was  used as linear algebra backend \cite{balay2016petsc}. The nonlinear problem was numerically solved through a SNES solver, where a secant predictor was used to identify the initial guess. %
At each step of the non-linear algorithm, the linear system was directly solved using MUMPS. The computational domain was discretized by subdividing each side of the rectangle into intervals of length $L/250$ and constructing a structured triangular mesh.%

\subsection{The geometrically linear case}
Suppose  that the deformation of our isotropic material is (geometrically) small but the material response is still  (physically) nonlinear. Let $E_1$ and $E_2$ be the eigenvalues of  the infinitesimal strain tensor
$
\tens{E}=\sym\nabla\vect{u} =(1/2)(\nabla\vect{u}+\nabla\vect{u}^T).
$
Then the strain energy density is $w(E_1,\,E_2)$.
In this  geometrical nonlinear approximation  the incompressibility constraint reduces to $\tr\tens{E}=\nabla\cdot\vect{u}=0$ and  therefore, without loss of generality, we canassume that  $0<E=E_1=-E_2$, and introduce the reduced energy density $\hat w(E)=w(E,\,-E)$.
We observe that
$
\hat{w}' = w_{,1}-W_{,2}$, and $ \hat{w}'' = w_{,11}+w_{,22}-2w_{,12}$, 
 where $w_{,j}= \partial w/\partial E_j$.  The Cauchy stress tensor is then  
$
\tens{S}=\partial w/\partial \tens{E}+p\tens{I}.
$

In the case of uniaxial traction,  the  homogeneous solution of the elastic equilibrium problem  is   $\vect{u}^{(0)} = E X_1 \vect{e}_1 -E X_2\vect{e}_2$ and $p^{(0)}=-w_{,2}$. To study linear stability  of such a solution, we again 
we  expand the unknown functions in terms of the small parameter $\epsilon$. Thus, we can write   $\tens{S} = \sum_j\epsilon^j \tens{S}^{(j)}$, where 
$
\tens{S}^{(1)} = \mathbf{C}^{(1)}:\tens{E}^{(1)} + p^{(1)}\tens{I}$, while  $\tens{E}^{(j)} = \sym\nabla\vect{u}^{(j)},$ and $\mathbf{C}^{(1)} = \left.\frac{\partial^2 W}{\partial\tens{E}\partial\tens{E}}\right|_{\tens{E}^{(0)}}.$
Using the results of \cite{Chadwick_1971} we  obtain that 
$C^{(1)}_{iijj}=w_{,ij},$ and 
 $C^{(1)}_{ijij}=\frac{1}{2}\frac{w_{,i}-w_{,j}}{e_i-e_j}$ for $i\neq j$,
while all the other components of the tensor $\mathbf{C}^{(1)}$ are equal to zero.
The linearized form of the balance of the linear momentum and of the incompressibility constraint reads
\[
\nabla\cdot\tens{S}^{(1)}=0\qquad \tr\tens{E}^{(1)}=0.
\]
Using again the stream function $\chi$, we can write  $\vect{u}^{(1)} = \partial\chi/\partial x_2\vect{e}_1 - \partial\chi/\partial x_1 \vect{e}_2$ where now 
$\chi = \imath A g(\gamma x_2)e^{\imath \gamma x_1}/\gamma$. The function $g(\gamma x_2)$  is a solution of the ODE 
\[
a g''''(\gamma  x_2)-2 b g''(\gamma  x_2)+c g(\gamma  x_2) = 0,
\]
where
$
a =c = \frac{\hat w'}{2E},$ and
$b =w''-\frac{\hat w'}{2E}$. The corresponding characteristic equation takes the form 
\[
\left(\alpha ^2+1\right) \left(\beta ^2+1\right) \left(\alpha  \beta  \left(\alpha ^2+\beta ^2+2\right) (\cosh (\alpha  \gamma  H) \cosh (\beta  \gamma  H)-1)-\left(\alpha ^2 \left(2 \beta ^2+1\right)+\beta ^2\right) \sinh (\alpha  \gamma  H) \sinh (\beta  \gamma  H)\right)=0
\]
where $\alpha$ and $\beta$ are defined as in \eqref{eq:ab} while the  solution  $g(\gamma x_2)$ has the same form as  \eqref{eq:g}.

We observe that the load maximum condition corresponds to a strain $E_\text{lm}$ such that $w''(E_\text{lm})=0$,representing a necessary condition for a bifurcation to occur \cite{HH75}. Similarly a sufficient condition for the instability of any wavenumber is provided by the condition of loss of strong ellipticity, which is reached at $b = -\sqrt{ac}$ \cite{OgdenBook}, i.e. when $E=E_\text{sm}$  defined again by the same condition  $w''(E_\text{se})=0$.
Therefore,  a bifurcation occurs if and only if  $E=E_\text{lm}=E_\text{se}$ for any value of $\gamma$, and all the wavenumbers become  unstable simultaneously.